\documentclass[12pt]{article}
\usepackage[utf8]{inputenc}
\usepackage[T1]{fontenc}

\usepackage[compress]{cite}
\usepackage{IEEEtrantools,stackengine}
\stackMath

\usepackage{authblk}
\usepackage{fullpage}
\usepackage{amssymb,amsmath}
\usepackage{adjustbox}
\usepackage{graphicx}
\usepackage{longtable}
\usepackage{xcolor}

\usepackage{siunitx}
\usepackage{csquotes}

\usepackage{xspace}

\usepackage{booktabs}
\usepackage{longtable}
\usepackage{multirow}

\usepackage[dvipsnames, table]{xcolor}

\usepackage[left]{lineno}

\definecolor{darkgreen}{rgb}{0.0, 0.5, 0.0}

\definecolor{lightyellow}{HTML}{FFE699}
\definecolor{red_revision}{HTML}{FF0000}

\usepackage{setspace}
\usepackage[unicode=true]{hyperref}
\hypersetup{breaklinks=true,
            pdfborder={0 0 0},
            allcolors=black}
\usepackage[%
  sort&compress
]{cleveref}
  \Crefname{appendix}{Supplement}{Supplements}
  \Crefname{figure}{Fig.}{Fig.}

\usepackage{times}
\usepackage{enumerate}
\usepackage{enumitem}
\usepackage{float}

\usepackage{subfig}
\usepackage{graphicx}
\usepackage[most]{tcolorbox}

\graphicspath{{plots/Plots/}}

\usepackage{rotating}
\usepackage{tabularx}
\usepackage{dcolumn}
\usepackage{pdflscape}
\usepackage{rotating}

\usepackage{array}

\usepackage{titlesec}

\titleformat{\subsubsection}
  {\normalfont\itshape}{\thesubsubsection}{1em}{}

\titleformat{\subsection}
  {\normalfont\bfseries}{\thesubsection}{1em}{}

\makeatletter
\renewcommand{\fps@figure}{H}         % default {tbp}
\renewcommand{\fps@table}{H}         % default {tbp}
\makeatother 

\renewcommand{\arraystretch}{1.2}

\allowdisplaybreaks
\usepackage{eurosym}
\usepackage{comment}

\usepackage{ntheorem}
\theoremseparator{:}

\usepackage{makecell}

\usepackage{changepage} % kommt in die Präambel, falls noch nicht vorhanden
\usepackage[table,xcdraw]
{xcolor}
\usepackage{tabularx}
\usepackage{ragged2e}

\begin{document}

\newpage

%%%%%%%%%%%%%%%%%%%%%%%%%%%%%%%%%%%%%%%%%%%%%%%%%%%%%%%%%%%%%%%%%%%%%%%%%%%%%%
% Title page

\title{\centering\LARGE\singlespacing A Meta-Analysis of the Persuasive Power of Large Language Models}

\renewcommand\Affilfont{\fontsize{9}{10.8}\selectfont}

% Author names
\author[1]{Lukas Hölbling\thanks{These authors contributed equally to this work.}}
\author[1,2]{Sebastian Maier\protect\footnotemark[1]}
\author[1,2]{Stefan Feuerriegel\thanks{Corresponding author: feuerriegel@lmu.de}}
% \thanks{Correspondence: feuerriegel@lmu.de}

\affil[1]{LMU Munich, Munich, Germany}
\affil[2]{Munich Center for Machine Learning (MCML), Munich, Germany}

\date{}

\maketitle

\newpage

\begin{abstract}\normalfont
\noindent
Large language models (LLMs) are increasingly used for persuasion, such as in political communication and marketing, where they affect how people think, choose, and act. Yet, empirical findings on the effectiveness of LLMs in persuasion compared to humans remain inconsistent. The aim of this study was to systematically review and meta-analytically assess whether LLMs differ from humans in persuasive effectiveness, and under which contextual conditions LLMs are particularly effective. We identified $7$ studies with 17,422 participants primarily recruited from English-speaking countries and $12$ effect size estimates. Egger’s test indicated potential small-study effects ($p = .018$), but the trim-and-fill analysis did not impute any missing studies, suggesting a low risk of publication bias. We then compute the standardized effect sizes based on Hedges' $g$. The results show no significant overall difference in persuasive performance between LLMs and humans ($g = 0.02$, $p = .530$). However, we observe substantial heterogeneity across studies ($I^2 = 75.97\%$), suggesting that persuasiveness strongly depends on contextual factors. In separate exploratory moderator analyses, no individual factor (e.g., LLM model, conversation design, or domain) reached statistical significance, which may be due to the limited number of studies. When considered jointly in a combined model, these factors explained a large proportion of the between-study variance ($R^2 = 81.93\%$), and residual heterogeneity is low ($I^2 = 35.51\%$). Although based on a small number of studies, this suggests that differences in LLM model, conversation design, and domain are important contextual factors in shaping persuasive performance, and that single-factor tests may understate their influence. Our results highlight that LLMs can match human performance in persuasion, but their success depends strongly on how they are implemented and embedded in communication contexts. 
\end{abstract}

\flushbottom
\maketitle
\thispagestyle{empty}

%\begin{center}
%\begin{tabular}{p{14.5cm}}
%\small
%\noindent\textbf{Keywords}: keyword, ...
%\end{tabular}
%\end{center}

\sloppy
\raggedbottom

%%%%%%%%%%%%%%%%%%%%%%%%%%%%%%%%%%%%%%%%%%%%%%%%%%%%%%%%%%%%%%%%%%%%%%%%%%%%%%
\newpage
\section{Introduction}
\label{sec:Introduction}

%Motivation: Large Language Models (LLMs) are increasingly deployed to persuade humans in marketing, politics, and education.

%Problem: However, it remains unclear whether LLMs are truly effective at persuasion when compared to human communicators.
    %ggf auch die negativen finding mal exemplarisch erwähnen 
%(combine both into one)

Large language models (LLMs) such as GPT-4 are increasingly embedded in communication settings that aim to shape attitudes, preferences, or behaviors \cite{Salvi2025, Bai2025, Matz2024}. For instance, marketing professionals use LLMs to generate persuasive product descriptions or targeted advertisements \cite{ghosh2024}. Political campaigns have experimented with AI-generated messaging to mobilize voters or improve communication styles \cite{Foos2024, Carrella2025}. In healthcare, LLMs are increasingly used for nudging healthier choices through personalized recommendations and framing techniques \cite{Chiam2024}, but their use is now also being explored in high-stakes clinical decision support, including assistance with disease diagnosis, triage, and treatment recommendations with potentially wide-ranging implications for patient care \cite{williams2024, thirunavukarasu2023, spitzer2025}.

%implications

These applications demonstrate the growing role of LLMs in persuasive communication, raising both promise and concern. On the one hand, LLMs offer scalable and adaptable tools for tailoring messages to individual recipients \cite{Matz2024}, potentially improving the effectiveness of public information \cite{Bai2025} or customer engagement \cite{Hartmann2025}, facilitating learning through multimodal and personalized educational support\cite{Bewersdorff2025}, and supporting more constructive civic discourse \cite{Tornberg2023}. On the other hand, the same technologies may be leveraged for manipulation, misinformation, or undue influence \cite{Feuerriegel2023, Bashardoust2024, Spitale2023, Goldstein2024, Geissler2025}, especially as LLMs become increasingly human-like in tone, reasoning, and responsiveness \cite{Bozdag2025}. Given this dual potential, a key question emerges: how persuasive are LLMs?

% current effect sizes 

Despite a growing body of empirical studies, the evidence on the persuasive capabilities of LLMs remains inconsistent \cite{Hackenburg2024, Teeny2024}. Reported effect sizes vary substantially: while some studies report that LLM-generated messages perform on par with or even better than human-written content \cite{Salvi2025, Schoenegger2025, Bai2025}, others find the opposite, showing that LLMs fail to outperform human communicators in direct comparison \cite{Chu2024, Lim2024}. However, we show later that these findings often stem from fundamentally different setups, including differences in model version, task format, and evaluation metric, which makes results rarely comparable across studies. As a result, there is no clear evidence for how persuasive LLMs are relative to humans. Here, we thus aim to answer the following research question: \emph{How effective are large language models at persuading humans, compared to human persuaders?}
%Empirical ambiguity: Prior studies report mixed findings, with some showing superior LLM performance and others favoring humans.
%(but oftenvery different setting due to different model, task etc , so not really comparable) 
  %TODO{Find relevant sources, supporting empirical ambiguity}
% with some studies favoring LLMs and others reporting superior human performance. 

% Therefore, we aim to address the following research question: 
% \begin{itemize}
%     \item \textit{How effective are large language models at persuading humans, compared to human persuaders?}
% \end{itemize}

% Methods: Brief summary how we try to answer the above question

%\TODO{ggf. eher past tense (methods sind ja auch in Past tense geschrieben}

To empirically answer this question, we conducted a systematic literature review and meta-analysis to assess the persuasive effectiveness of LLMs. We identified $n=7$ eligible studies ($m = 17,422$ participants) from an initial pool of $n=112$ screened records. We adopted highly stringent inclusion criteria, which reduced the number of eligible studies but ensured that effects were not driven by software design or other confounding factors, thereby isolating the persuasive impact of the LLMs themselves. For each study, we extracted standardized effect sizes and computed pooled estimates based on Hedges'$g$, which allowed us to quantify the persuasive performance of LLMs relative to human communicators. In addition, we performed exploratory moderator analyses to assess whether differences in model type (e.g., GPT-4 vs. Claude 3.7), interaction format (e.g., interactive dialogues vs. one-shot messages), or domain (e.g., health vs. politics) explain variation in persuasive performance. The moderator analyses revealed that the observed heterogeneity in persuasive outcomes is, to a large extent, explained by contextual factors, which offer new insights into when and how LLMs are more or less persuasive compared to humans. By quantitatively integrating results across studies, our work provides the first systematic meta-analytic synthesis of LLM persuasiveness relative to humans and the conditions under which LLMs are most effective.
%Outline: Data collection, statistics, heterogeneity, discussion

% The remainder of this paper is structured as follows. Section \ref{sec:theoretical_background} outlines relevant theoretical perspectives and recent research on persuasive LLMs to contextualize this study. Section \ref{sec:methods} outlines the literature search, data extraction, and statistical procedures used in this meta-analysis. Section \ref{sec:results} presents the empirical findings of the meta-analysis, including moderator analyses to explain observed variation. Section \ref{sec:discussion} discusses the implications of these findings for theory and practice. Section \ref{sec:conclusion} concludes by summarizing the key contributions and suggesting directions for future research.

\newpage
\section{Theoretical background}
\label{sec:theoretical_background}

%Können wir den ersten absatz etwas trennen (bzw wir brauchen hier eine tws generell estrukutrI): mehr in (1) was ist persuasion als behavioral construct oder (2) wie kann es operationalisiert (gemessen werden)  

Persuasion refers to the deliberate attempt to influence others' beliefs, attitudes, or behavioral intentions through communication \cite{Perloff1993}. In social psychology, persuasion is considered a foundational mechanism for shaping human behavior and decision-making, with extensive research in domains such as health messaging, political advocacy, and consumer communication \cite{PettyCacioppo1986,CranoPrislin2006}. 

Prominent theoretical models, including the Elaboration Likelihood Model (ELM), posit that persuasion operates via two distinct cognitive routes: a central route involving deliberate elaboration of message arguments and a peripheral route relying on heuristics such as source credibility or affective cues\cite{PettyCacioppo1986}. The likelihood of central versus peripheral processing depends on the recipient's motivation and ability to elaborate, which, in turn, are shaped by factors such as personal relevance, prior knowledge, and contextual complexity \cite{Cacioppo1996}. The Persuasion Knowledge Model (PKM) further explains that, when recipients recognize a message as an attempt to persuade, their reactions may change accordingly, either by resisting or reinterpreting the message -- which often activates the central processing route. For example, on social media, knowing an influencer is paid for a recommendation can lead followers to question the message (central route), whereas unawareness may prompt reliance on cues like attractiveness or popularity (peripheral route). While the dual-process model explains how people process persuasive messages, the model does not stipulate how processing translates into intentions and behavior. 

Several theories may offer insights into how persuasion varies across individuals. The Theory of Planned Behavior \cite{Ajzen1991} emphasizes the role of attitudes, perceived norms, and behavioral control in shaping intentions and actions. Applied to the influencer example from above, the likelihood that a follower will buy a promoted product increases when, for example, followers view the product positively, believe that important others approve, and feel able to purchase and use it.

Together, these theoretical models highlight that persuasive effects do not stem from the message content alone, but emerge from dynamic interactions between message, communicator, and recipient. While originally developed to explain human-to-human persuasion, these theories are now increasingly used to assess whether AI-generated content—such as that produced by large language models—can trigger similar shifts in beliefs, attitudes, and behavioral intentions \cite{Bai2025}.

Empirical studies in persuasion typically assess effectiveness through one or more outcome types: (i)~changes in attitude (later referred to as ``attitude''), (ii)~shifts in behavioral intention  (later referred to as ``intention''), and (iii)~actual behavior (later referred to as ``behavior'')\cite{OKeefe2018,CranoPrislin2006}. These outcomes are often measured using Likert-scale items or behavioral indicators such as agreement rates, policy support, or compliance decisions. Another common approach is to use perceived message effectiveness (PME) as a proxy for actual persuasion. In some cases, researchers aggregate multiple outcomes into composite indices to capture a broader construct of persuasive impact \cite{Bai2025,Karinshak2023,Chu2024,TimmJasper2025}. Of note, understanding how persuasion is operationalized is crucial later for interpreting and comparing effect sizes across studies. These measures vary not only in their scale and format but also in the underlying psychological construct they reflect, ranging from cognitive evaluations of message quality to affective or behavioral responses. These differences in outcome measures may account for part of the variability in effect sizes reported in the literature and thus inform our selection of moderators in our meta-analytic analysis.

\newpage
\section{Method}
\label{sec:methods}

%Sentence introducing the next chapter
%In this section, we describe the procedures used to identify relevant studies and outline the statistical methods applied.

\subsection*{Search strategy}
%PRISMA Framework, searched the Databases: WoS, ArXiv, ACM, SSRN, OSF/ included Peer-reviewed & Preprints due to novelty of research/ only included english paper after Dec. 2022 (introduction of GPT-3) 

Our data collection follows the PRISMA 2020 framework \cite{PRISMA2020} for systematic reviews. We searched five databases widely used in empirical artificial intelligence research: Web of Science, ACM Digital Library, arXiv, SSRN, and OSF. This selection covers both peer-reviewed journal articles from fields such as general science and psychology, and preprints from emerging research areas. The inclusion of preprints was intentional, given the rapid development of large language models and the recency of this research field. To focus only on LLMs and not traditional machine learning methods, we restricted our search to English-language papers published after December 2022, as this marks the point at which GPT-3 became publicly available. The knowledge cutoff of our search was May~22,~2025.

%State search query and that it was inspired by Holzner et al.(2025) and Schemmer et al. (2022)
We adopted a stringent search strategy that required a direct experimental comparison between LLMs and human communicators to ensure that any observed differences in persuasiveness reflect the LLMs’ capabilities rather than artifacts arising from software design, contextual framing, or other confounding factors. This, in turn, allows us to assess the persuasive impact of LLMs alone.  This deliberately narrow focus improves internal validity and ensures greater comparability across measures, thereby raising data quality.

We intentionally designed our query to capture empirical studies on LLM-based persuasion. The formulation is inspired by prior research on analyzing the effects of AI on human behavior  \cite{Holzner2025,Schemmer2022}, but was adapted to the context of our research question: 

\begin{tcolorbox}[colback=gray!5!white, colframe=black!75, title=Search query, fonttitle=\bfseries]
\ttfamily\footnotesize\singlespacing
\vspace{-0.4cm}
TITLE("persuasion" OR "persuasive" OR "persuade" OR "persuasiveness") \\
AND ("LLM" OR "Large Language Model" OR "AI" OR "artificial intelligence" OR "ChatGPT" OR "GPT")
\end{tcolorbox}

%Explain the literature review process (Brocke et al. as source) in detail, inlcuding reasons for exclusion/ inclusion (Studies had to include a direct comparison of LLM- and human-generated messages with quantitative persuasion outcomes), and number of exclusion/inclusions at every stage

\begin{figure}[H]
\centering
\includegraphics[
  trim=1cm 12cm 4.5cm 2cm, % links unten rechts oben (in cm)
  clip,
  width=\linewidth
]{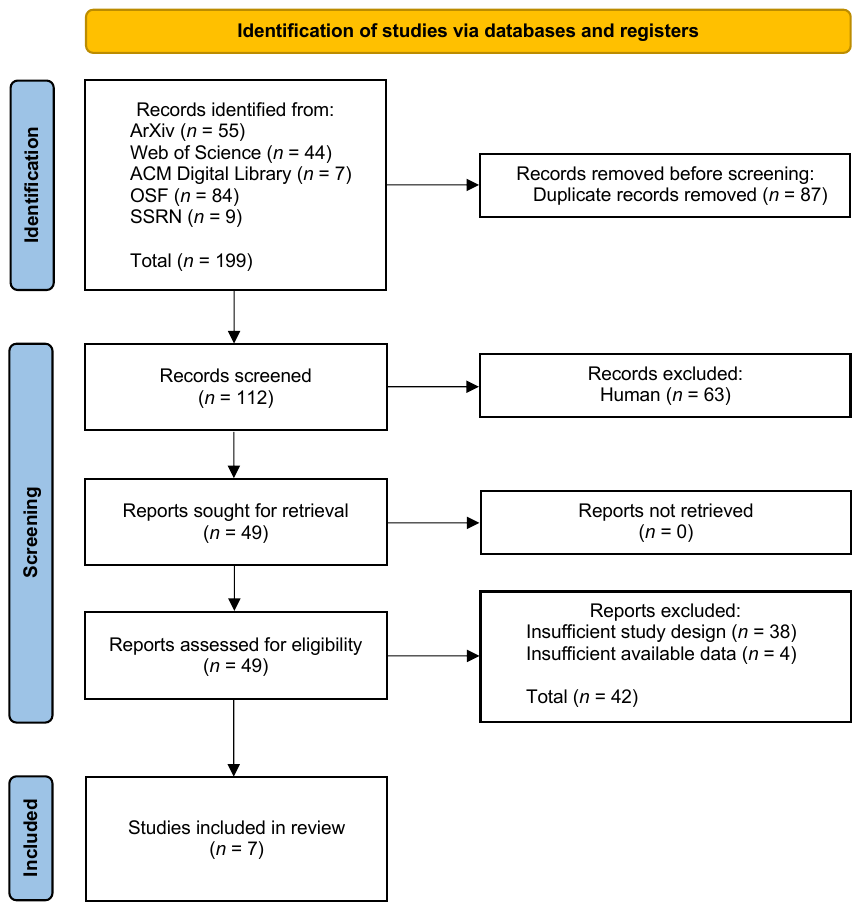}
\caption{\textbf{Flow diagram for study selection.} The diagram summarizes the number of records identified, screened, and included in the meta-analysis following the PRISMA 2020 guidelines.}
\label{fig:prisma}
\end{figure}

Screening and eligibility assessment were conducted by the first author. Our procedure follows a structured approach \cite{Brocke2009}, which emphasizes methodological transparency through the systematic documentation of inclusion and exclusion decisions at each stage of the review. The process of study selection is summarized in Figure~\ref{fig:prisma}, following the format of the PRISMA 2020 flowchart \cite{PRISMA2020}. Initially, we retrieved a total of $n = 199$ records through our database search. After removing duplicates ($n = 87$ duplicates); $n = 112$ unique records remained. Titles and abstracts were screened for relevance, based on which we excluded $n = 63$ publications that clearly did not meet the inclusion criteria (e.g., theoretical essays or no human comparison).

We then assessed $n = 49$ full-text articles for eligibility. Studies were included if they met the following criteria:
(i)~the study provided a direct comparison of persuasive performance between LLM-generated and human-generated content,
(ii)~the study design allowed for independent observations (i.e., between-subjects comparison at the message or participant level),
(iii)~the outcome measure reflected a form of persuasive effectiveness (e.g., attitude change, compliance, policy support, or perceived message effectiveness), and
(iv)~the study reported sufficient statistical information (e.g., means and standard deviations) to compute standardized effect sizes. During this stage, studies were excluded either due to an insufficient study design (e.g., lack of a valid comparison between LLMs and human messages) or because key statistical information required for effect size computation was not available. 

As a result, $n = 38$ studies were removed for methodological reasons, and $n = 4$ due to missing outcome data. In cases of uncertainty, inclusion decisions were discussed and resolved by the first author in consultation with the second author.
%n authors were contaced manually to adress the issue of insufficient statistics
For four papers with missing statistical information, we contacted the corresponding authors via email to request the necessary data. However, we did not receive sufficient information, and all four studies were eventually excluded from the analysis.

In total, we identified $n = 7$ studies that fulfilled all eligibility criteria and were included in the meta-analysis. Some of these studies contained multiple experimental conditions or reported more than one relevant persuasion outcome. Accordingly, the final dataset comprises $12$ independent effect size estimates.

\subsection*{Data collection}

%State and Explain what and why of data/variables we extracted of n remaining papers

All $n = 7$ studies that passed the full-text screening were systematically coded into a structured dataset. For each study, we extracted: (i)~metadata (e.g., title, author, publication year), (ii)~characteristics of the study design (e.g., study year, country, study environment, publication status), and (iii)~experimental features such as the LLM model, conversation design (e.g., interactive vs. one-shot), domain (e.g., health, politics), and content length (e.g., words $<$ 200; multiple round dialogue). In addition, we recorded variables related to personalization, participant type, recruitment channel, and outcome definitions. The variables were selected to allow for a comprehensive characterization of the literature and to enable moderator analyses. The full list is in Table~\ref{tab:dimensions}. The collected data is publicly available in our GitHub repository (see Data Availability statement).

%\TODO{IMHO past das nicht zur zeile "outcome typoe" -- hier sind andere Namen also dort, }
To assess persuasive effectiveness across studies, we proceeded as follows. We recorded the specific outcome measures used in each experiment (see Table~\ref{tab:meta_analysis_literature}). Across studies, persuasion was operationalized through a broad range of constructs, including attitude change, behavioral intention, compliance rate, perceived message effectiveness (PME), and policy support serve as proxies for underlying belief change \cite{OKeefe2018, CranoPrislin2006}. While all these indicators aim to capture persuasion, they may differ in their proximity to actual behavioral change \cite{Albarracin2018}. For example, perceived message effectiveness is often treated as a proxy for real-world impact but may overestimate persuasive success due to social desirability or hypothetical framing \cite{Dillard2007}. Conversely, compliance rates or expressed intention may provide more tangible evidence of influence \cite{Sheeran2002}. To address this heterogeneity, we selected, for each study, the outcome measure that most closely reflected actual changes in belief or behavior. Specifically, we prioritized behavioral intentions or compliance-based outcomes over purely perceptual indicators such as message effectiveness, unless only the latter were available.

\begin{table}[H]
\centering
\singlespacing
\footnotesize
\vspace{-0.5cm}
\begin{tabularx}{\linewidth}{lX}
\toprule
\textbf{Dimensions} & \textbf{Values} \\
\midrule
LLM model & GPT-3.x (i.e., GPT 3, GPT 3-davinci, GPT 3.5), GPT-4.x (i.e., GPT 4, GPT 4o mini), Claude 3.x (i.e., Claude Sonnet 3.5, Claude Sonnet 3.7) \\
\hline
Interaction type & one-shot, interactive\\
\hline
Domain & politics, health, other (propaganda, quiz, mixed) \\
\hline
Content type & words < 200, 200 < words < 500, multiple round dialogue\\
\hline
Participant type & general population, educated \\
\hline
Personalization & yes, no\\
\hline
Recruitment source & Prolific, mTurk, Lucid\\
\hline
Outcome type & attitude, intention, behavior \\
\hline
Outcome variable & attitude change, compliance rate, policy support, behavioral intention, percent agreement, perceived message effectiveness (PME) \\
Outcome measurement & 7-point Likert scale, 5-point Likert scale, 101-point scale, binary proportion \\
\bottomrule
\end{tabularx}
\caption{\textbf{Extracted dimensions.} The table summarizes the dimensions coded from the included studies. 
Some dimensions were aggregated for analysis, specifically for the LLM model and domain. The interaction type “one-shot” refers to message conditions where the LLM produces a single static output, whereas “interactive” refers to multi-turn setups allowing adaptive communication.
}
\label{tab:dimensions}
\end{table}

\subsection*{Statistical analysis}

To standardize persuasive effects across studies, we computed effect sizes using Hedges' $g$. We aimed to extract reported values of Cohen's $d$ directly from the primary studies. However, as none of the included studies reported standardized effect sizes, we calculated Cohen's $d$ based on available summary statistics (e.g., means, standard deviations, sample sizes, $t$- or $F$-values), following established conversion formulas \cite{Borenstein2021,Dunlap1996,Rosnow1996}. All values were subsequently corrected for small-sample bias using the standard Hedges' $g$ correction \cite{Hedges1981}. To maintain statistical independence, only one effect size per experiment was included in the analysis. In addition, each estimate is reported alongside its $95\%$ confidence interval~(CI).

%Random-effects model: Explain that we chose a random-effects model and also calculated 95% CIs
We estimated a random-effects model to account for heterogeneity in LLM model, conversation design, domain, content length, and outcome measurement across studies. Here, we again followed Cochrane's approach \cite{Cochrane2019}, which reflects the assumption that effect sizes may vary systematically between settings. The pooled effect was computed using the restricted maximum-likelihood (REML) estimator \cite{Viechtbauer2005}, and a $95\%$ confidence interval was calculated to quantify the expected range of true effects.

%heterogeneity assessment: we estimated heterogeneity via τ² and I², and conducted subgroup and meta-regression analyses.
To assess heterogeneity across studies, we estimated between-study variance using the $\tau^2$ statistic and quantified residual variability via the $I^2$ statistic and Cochran's $Q$ test \cite{Borenstein2021}. To identify potential sources of heterogeneity, we performed subgroup analyses and meta-regressions based on predefined moderators. Specifically, we examined whether effect sizes differed depending on LLM model, conversation design, and domain.

%Bias assessment: We assessed bias using influence diagnostics, leave-one-out analyses, Egger's test, funnel plots, and the trim-and-fill method.
To examine potential publication bias, we applied Egger's regression test \cite{Egger1997} and the trim-and-fill procedure \cite{TrimmFill2000}, following established recommendations for meta-analyses with at least ten effect sizes \cite{Cochrane2019}. In addition, we computed influence diagnostics and leave-one-out analyses to evaluate the robustness of individual studies on the pooled effect estimate \cite{InfluenceDiagnostic2010}.

%maybe one sentence about which R version was used and that everything can be found in GitHub
All analyses were conducted in R (version 4.5.0) using the \texttt{metafor} package (version 4.8-0). The complete analysis code, research data, and output are available on GitHub (see Data Availability statement).

%All numerical results (e.g., effect sizes, confidence intervals, p-values) were rounded to three decimal places for consistency. 

\newpage

\begin{landscape}

\begin{table}[H]

  \centering
%  \vspace*{\fill}
%  \begin{adjustbox}{angle=90,center}
    
    \fontsize{5}{5.8}\selectfont
    \setlength{\tabcolsep}{1pt}
    \renewcommand{\arraystretch}{1.0}
    \rowcolors{2}{lightgray!40}{white}  % Alternating row color: gray & white
    \begin{tabular}{l|l|l|l|l|l|l|l|l|l|l|l|l|r}
\rowcolor{gray!30}
\hline
Study\_id & Report\_id & Title & LLM model & Conversation Design & Domain & Content length & Personalization & Recruitment source & Outcome type & Outcome variable & Outcome measurement & \#Participants ($n$) \\
\hline\hline
S\_1 & R\_1 & Large Language Models Are More Persuasive Than Incentivized Human Persuaders \cite{Schoenegger2025} & Claude Sonnet 3.5 & interactive & quiz & multiple round dialogue & NO & Prolific & behavior & compliance rate & binary proportion & 708 \\
S\_2 & R\_1 & A Framework to Assess the Persuasion Risks Large Language Model Chatbots Pose to Democratic Societies \cite{Chen2025} & Claude Sonnet 3.7 & interactive & politics & multiple round dialogue & NO & Prolific & attitude & attitude change & 5-point likert scale & 3900 \\
S\_2 & R\_2 & A Framework to Assess the Persuasion Risks Large Language Model Chatbots Pose to Democratic Societies \cite{Chen2025} & Claude Sonnet 3.7 & interactive & politics & multiple round dialogue & NO & Prolific & attitude & attitude change & 5-point likert scale & 4511 \\
S\_3 & R\_1 & On the Conversational Persuasiveness of GPT-4 \cite{Salvi2025} & GPT 4 & interactive & politics & multiple round dialogue & YES & Prolific & attitude & attitude change & 5-point likert scale & 300 \\
S\_3 & R\_2 & On the Conversational Persuasiveness of GPT-4 \cite{Salvi2025} & GPT 4 & interactive & politics & multiple round dialogue & NO & Prolific & attitude & attitude change & 5-point likert scale & 300 \\
S\_4 & R\_1 & Can AI tell good stories? Narrative transportation and persuasion with ChatGPT \cite{Chu2024} & GPT 4 & one-shot & health & 200 < words < 500 & NO & Prolific & attitude & policy support & 5-point likert scale & 348\\
S\_4 & R\_2 & Can AI tell good stories? Narrative transportation and persuasion with ChatGPT \cite{Chu2024} & GPT 4 & one-shot & health & 200 < words < 500 & NO & Prolific & intention & behavioral intention & 5-point likert scale & 386\\
S\_5 & R\_1 & How persuasive is AI-generated propaganda? \cite{Goldstein2024} & GPT 3-davinci & one-shot & propaganda & words < 200 & NO & Lucid & attitude & scaled agreement & 5-point likert scale & 4116\\
S\_6 & R\_1 & Working With AI to Persuade: Examining a Large Language Model's Ability to Generate Pro-Vaccination Messages \cite{Karinshak2023} & GPT 3 & one-shot & health & words < 200 & NO & mTurk & attitude & PME & 5-point likert scale & 802\\
S\_6 & R\_2 & Working With AI to Persuade: Examining a Large Language Model's Ability to Generate Pro-Vaccination Messages \cite{Karinshak2023} & GPT 3 & one-shot & health & words < 200 & NO & mTurk & attitude & PME & 5-point likert scale & 443\\
S\_7 & R\_1 & LLM-generated messages can persuade humans on policy issues \cite{Bai2025} & GPT 3 & one-shot & politics & words < 200 & NO & Prolific & attitude & policy support & 101-point scale & 600\\
S\_7 & R\_2 & LLM-generated messages can persuade humans on policy issues \cite{Bai2025} & GPT 3.5 & one-shot & politics & words < 200 & NO & Prolific & attitude & policy support & 101-point scale & 1008\\
\hline
\end{tabular}
%  \end{adjustbox}
%  \vspace*{\fill}
  \caption{\textbf{Overview of included studies.} 
The table summarizes all experimental comparisons included in our meta-analysis. Each row represents one effect size estimate, uniquely identified by a combination of \texttt{Study\_ID} and \texttt{Report\_ID}. The columns provide information on the study title, LLM model, conversation design, domain, content length, presence of personalization, recruitment source, outcome type, outcome variable, outcome measurement, and number of participants. Participant samples were predominantly recruited from the United States.}
  \label{tab:meta_analysis_literature}
\end{table}
\end{landscape}

\newpage
\newpage
\section{Results}

%Sentence introducing the next chapter

This section presents the results of the meta-analysis on the persuasive effectiveness of large language models (LLMs) compared to human communicators. We first summarize descriptive characteristics of the included studies, followed by an analysis of the overall effect. To assess potential sources of heterogeneity, we then examine the role of study domain, LLM model, and interaction format.

\subsection*{Descriptive summary}

%Explain the difference/ key characteristics of included studies, i.e. different domains, LLM models, interaction type, outcome type
The final dataset comprised $12$ independent effect size estimates derived from $7$ unique publications, each comparing the persuasive effectiveness of LLM-generated and human-generated content. While all studies adhered to the core design of contrasting messages produced by artificial and human communicators, they varied substantially in their experimental setup. Note that the relatively small sample size is intentional and reflects our stringent inclusion criteria, which were designed to isolate the direct persuasive impact of LLMs while excluding studies that primarily varied confounding factors such as software design or interface features.
The included studies employed different LLMs, ranging from early versions such as GPT-3 ($n = 3$) and GPT-3.5 ($n = 1$) to more advanced models like GPT-4 ($n = 4$) and Claude Sonnet ($n = 3$). 
In terms of conversation design, the majority of studies employed a one-shot communication format ($n = 7$), in which participants evaluated a single static message. Comparatively fewer studies implemented interactive exchanges ($n = 5$), in which humans and LLMs engaged in cooperative, back-and-forth communication. This format allowed the LLM to adapt its responses dynamically based on human input. This distinction is relevant, as prior research suggests that interactive communication can enhance persuasion \cite{Sundar2005,Sundar2015}.
Also, the studies spanned multiple domains, including politics ($n = 6$), health ($n = 4$), and other areas such as mixed domains, quizzes, and propaganda ($n = 2$ in total).

\subsection*{LLM vs. human persuasion}

\begin{figure}
\begin{center}
\includegraphics[width=.6\linewidth]{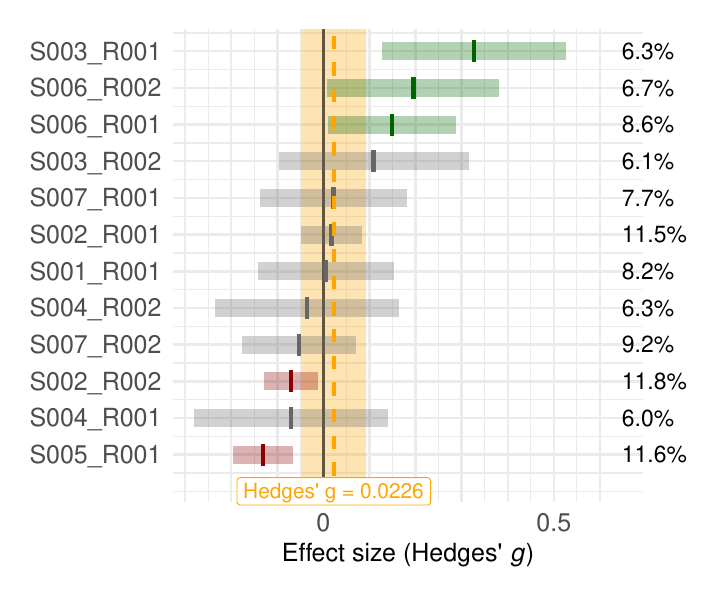}
\end{center}
\caption{\textbf{Forest plot of effect sizes (Hedges' $g$) comparing the persuasive effects of LLMs vs humans.} Each line represents one effect size estimate with its $95\%$ confidence interval. The orange shaded area indicates the $95\%$ confidence interval of the pooled effect. Study weights are shown on the right. Positive values indicate that LLMs were more persuasive than humans, whereas negative values indicate that humans were more persuasive. The overall effect size was very small and non-significant ($g = 0.02$, $p = .530$, $95\%$ CI [$-0.048$, $0.093$]). Substantial heterogeneity was observed ($I^2 = 75.97\%$).}
\label{fig:forest_plot_main}
\end{figure}

%The meta-analysis reveals a negligible and statistically non-significant difference in persuasive effectiveness between LLMs and humans.
To address our main research question, namely, whether LLMs are more effective at persuading humans compared to human persuaders, we conducted a random effects meta-analysis on all eligible studies that directly compared LLM-generated and human-generated persuasive messages. The corresponding forest plot is presented in Figure~\ref{fig:forest_plot_main}. The meta-analysis reveals a negligible and statistically non-significant difference in persuasive effectiveness between the two groups. The pooled effect size is Hedges' $g = 0.02$ ($p = .530$), with a $95\%$ confidence interval of [$-0.048$, $0.093$].

%Interpretation: On average, LLMs and human communicators are equally persuasive. However, this does not preclude important differences in subgroups or study settings.  Over half of the total variance is due to real differences between studies, not sampling error. This supports the need to investigate moderators that might explain when and why LLMs perform differently.
On average, we find insufficient evidence for a difference in persuasiveness between LLMs and humans. This finding proved robust across both the full dataset and a sensitivity analysis restricted to peer-reviewed studies only, indicating that the inclusion of preprints does not bias our results (see Table~\ref{tab:peer_review_comparison}). However, this overall result does not rule out meaningful differences between specific study contexts or subgroups. Given that we found substantial heterogeneity ($I^2 = 75.97\%$), this suggests that over half of the observed variance in effect sizes reflects actual differences across studies rather than random sampling error. Motivated by this, we next exploratively examine potential moderators that may help explain under which conditions LLMs outperform or underperform relative to humans.

\subsection*{Heterogeneity}
%Introduction sentence introducing moderator analysis
To better understand the variation in persuasive effectiveness across studies, we performed an exploratory analysis to investigate whether differences in model type, conversation design, and domain may systematically moderate the observed effects. It is important to note that several of these analyses are based on small subgroups, and, thus, the statistical power of some analyses may have been insufficient to detect meaningful differences.

\subsubsection*{Moderating role of the LLM model}

%The specific LLM used (e.g., GPT-3.5 vs. GPT-4) does not significantly affect persuasive performance.
We examined whether the specific LLM  (e.g., GPT-3.5, Claude 3.7) used influences persuasive effectiveness. Due to the limited sample size, we aggregated the models ``GPT-3'' and ``GPT-3.5'' into a single category labeled ``GPT-3.x'' (N = 5), ``GPT-4'', ``GPT-4o'', and ``GPT-4o mini'' into ``GPT-4.x'' (N = 4), and ``Claude Sonnet 3.5'' and ``Claude Sonnet 3.7'' into ``Claude 3.x'' (N = 3) (Figure~\ref{fig:moderator_plots}). The overall model was not significant ($QM(2) = 1.04$, $p = .596$). None of the model contrasts reached statistical significance. Using GPT-4.x as the reference category, neither GPT-3.x ($b = -0.06$, $95\%$ CI [$-0.25, 0.12$], $p = .491$) nor Claude 3.x ($b = -0.10$, 95\% CI [$-0.30$, $0.10$], $p = .312$) showed significantly different persuasive effects. This suggests that no single LLM consistently outperforms others in terms of persuasive power. However, given the limited number of studies per model type and the potential for contextual dependencies (e.g., task framing, message type), these findings should be interpreted with caution.

\subsubsection*{Moderating role of the conversation design}

We further examined whether the degree of interactivity between LLM and the recipient affected persuasive effectiveness. Specifically, we compared (i)~one-shot message conditions, where the LLM produces a single static output (N = 8), with (ii)~interactive setups that allow for adaptive multi-turn communication (N = 4). The analysis revealed no statistically significant difference between formats ($QM(1) = 0.40$, $p = .525$). One-shot messages yielded a slightly lower effect size ($b = -0.048$, $95\%$ CI [$-0.200$, $0.101$], $p = .525$), but this difference was not significant. Despite the null finding, the direction of the effect suggests that interactive formats may offer a modest advantage (Figure~\ref{fig:moderator_plots}). Given that interactivity enables LLMs to respond to user input, this pattern may reflect potential opportunities for engagement or message adaptation.

\subsubsection*{Moderating role of the domain}

\begin{figure}[H]
\centering

\begin{minipage}{0.4\textwidth}
    \centering
    \includegraphics[width=\linewidth]{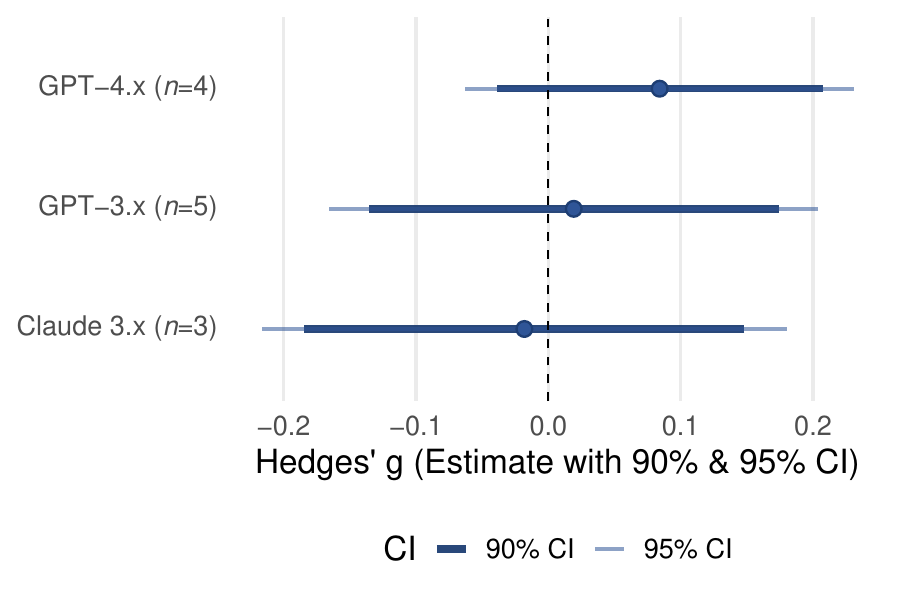}
    \caption*{\textbf{(a)} LLM model}
\end{minipage}

\vspace{0.5cm} % Abstand zwischen den Plots

\begin{minipage}{0.4\textwidth}
    \centering
    \includegraphics[width=\linewidth]{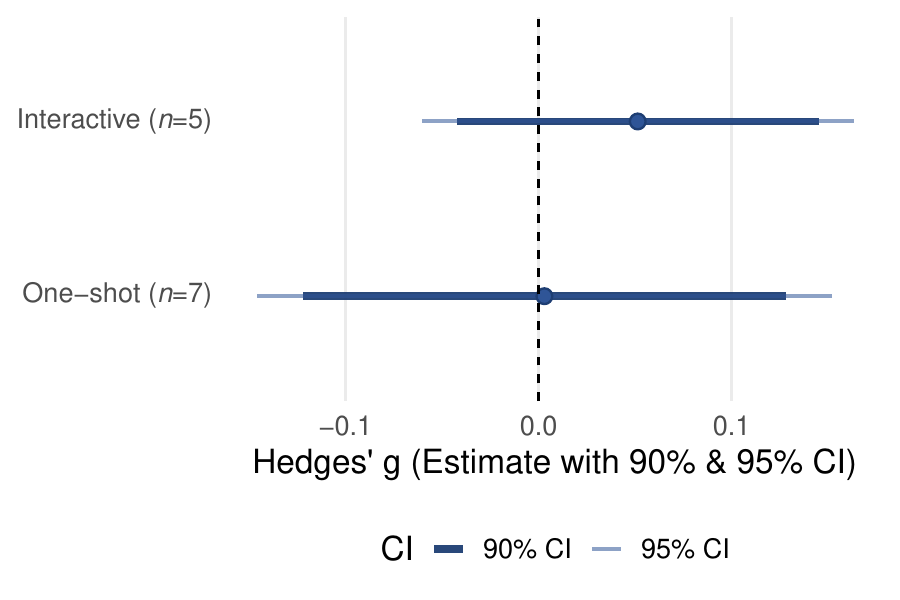}
    \caption*{\textbf{(b)} conversation design}
\end{minipage}

\vspace{0.5cm}

\begin{minipage}{0.4\textwidth}
    \centering
    \includegraphics[width=\linewidth]{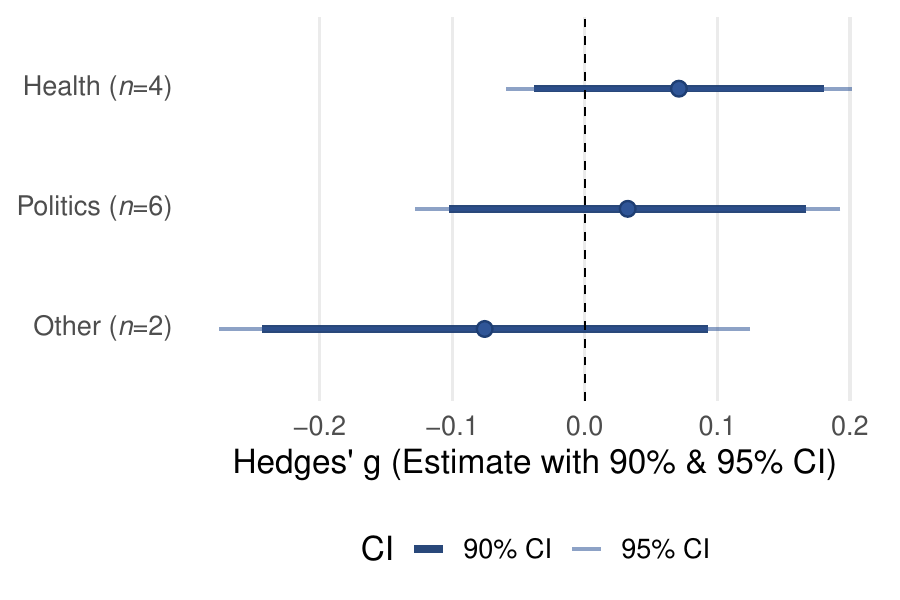}
    \caption*{\textbf{(c)} domain}
\end{minipage}

\caption{\textbf{Meta-regression estimates for moderator analyses.}  
Each plot shows the estimated difference in effect size ($g$) with $95\%$ (thin light blue bars) and $90\%$ (thick dark blue bars) CIs for (a)~LLM model, (b)~conversation design, and (c)~domain. The numbers in parentheses indicate the number of effect sizes ($n$) contributing to each estimate. Each moderator comparison is based on small subgroup sizes (see $n$ values in parentheses); results should thus be interpreted cautiously.}
\label{fig:moderator_plots}
\end{figure}

%The Domains Propaganda, Quiz & Mixed where grouped together as "Other" because k<2 (included studies) for each domain
We further examined whether the domain, such as politics or health, moderates the persuasive effectiveness of LLMs relative to human communicators (Figure~\ref{fig:moderator_plots}). Due to the limited sample size, we aggregated the domains ``Propaganda'', ``Quiz'', and ``Mixed'' into a single category labeled ``Other''. However, we did not see that the domain explained variation in effect sizes ($QM(2) = 2.17$, $p = .338$). For example, LLM persuasion in a political context was almost equally effective as in the health domain  ($b = -0.039$; $95\%$ CI:[$-0.199$, $0.122$], $p = .637$), whereas studies categorized as ``Other'' tended to show slightly less effective LLM persuasion ($b = -0.147$; $95\%$ CI:[$-0.347$, $0.054$], $p = .152$). Overall, these results do not provide clear evidence that the domain affects how persuasive LLMs are compared to humans. However, this may be due to limited data.

% --- Heterogeneity plots as one combined figure (a/b/c) via minipage ---

\subsubsection*{Combined moderator analysis}

To assess the overall impact of different factors, we included all moderators in a combined model (see Table~\ref{tab:combined_model}). This model explained a large share of the between-study variation ($R^2 = 81.93\%$) and reduced the remaining unexplained heterogeneity to a low level ($I^2 = 35.51\%$). When controlling for the other moderators simultaneously, several factors showed statistically significant independent effects on persuasive effectiveness: For example, Claude 3.x was less persuasive than GPT-4.x ($b = -0.236$, 95\% CI:[$-0.406$, $-0.065$], $p = .007$), one-shot message formats were less persuasive than interactive setups ($b = -0.494$, 95\% CI: [$-0.768$, $-0.220$], $p > .001$), and the domain category ‘Politics’ was associated with lower persuasive effectiveness compared to ‘Health’ ($b = -0.221$, 95\% CI: [$-0.384$, $-0.059$], $p = .008$). Notably, the comparison between GPT-3.x and GPT-4.x reversed when controlling for domain and interaction type: GPT-3.x showed significantly higher persuasive effectiveness than GPT-4.x ($b = 0.219$, 95\% CI: [$0.017$, $0.420$], $p = .033$), suggesting that the descriptive advantage of GPT-4.x in the univariate analysis might have been confounded by differences in study design or domain. These results highlight that LLMs can match human performance in persuasion, but these abilities may depend strongly on how they are implemented and embedded in communication contexts. However, these results should be interpreted with considerable caution, as the number of studies is very small relative to the number of parameters, and the observed drop in residual heterogeneity may therefore reflect overfitting.

\begin{table}[ht]
\centering
\footnotesize
\begin{tabular}{lrrrrrrl}
\toprule
Covariate & Estimate & SE & $z$ & $p$-value & Lower 95\% CI & Upper 95\% CI & \\
\midrule
Intercept & 0.442 & 0.115 & 3.855 & $<$0.001 & 0.217 & 0.667 & ***\\
Conversation Design: one-shot prompt & $-$0.494 & 0.140 & $-$3.538 & $<$0.001 & $-$0.768 & $-$0.220 & ***\\
Domain: Other & $-$0.271 & 0.081 & $-$3.342 & 0.001 & $-$0.429 & $-$0.112 & ***\\
Domain: Politics & $-$0.221 & 0.083 & $-$2.669 & 0.008 & $-$0.384 & $-$0.059 & **\\
LLM: GPT-3.x & 0.219 & 0.103 & 2.128 & 0.033 & 0.017 & 0.420 & *\\
LLM: Claude 3.x & $-$0.236 & 0.087 & $-$2.711 & 0.007 & $-$0.406 & $-$0.065 & **\\
\addlinespace
\bottomrule
\multicolumn{8}{l}{\footnotesize Significance level: $^{***}$\,$p < 0.001$; $^{**}$\,$p < 0.01$; $^{*}$\,$p < 0.05$}
\end{tabular}
\caption{Meta-regression results for the combined model. The model is based on $n = 12$ effect size estimates (Model fit: $R^2$ = 81.93\% (variance explained), $I^2$ = 35.51\% (residual heterogeneity))}
\label{tab:combined_model}
\end{table}

\begin{figure}[H]
\centering
\includegraphics[width=\linewidth]{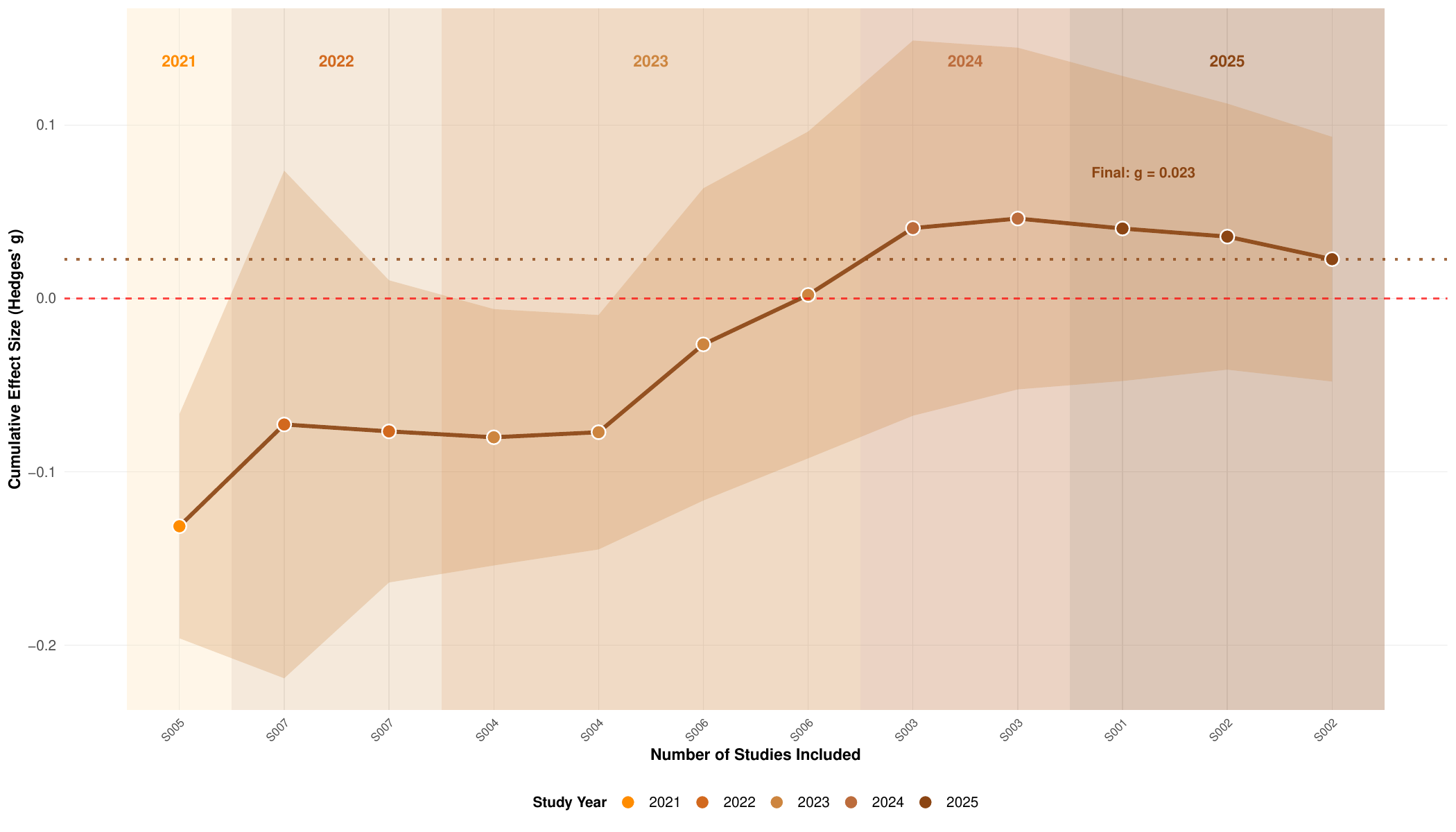}
\caption{\textbf{Cumulative meta-analysis of LLM persuasion effects by study year.} 
The plot shows how effect size estimates evolved as studies were added chronologically from 2021 through 2025. Points represent cumulative estimates at each step, colored by study year. The shaded ribbon indicates 95\% confidence intervals around the cumulative estimates. Background shading highlights different time periods. The horizontal dashed line marks zero effect, while the dotted line shows the final pooled estimate. Study IDs (S005, S007, etc.) on the x-axis indicate the sequential addition of studies ordered by study year.}
\label{fig:cumulative_evolution}
\end{figure}

\subsection*{Time trend}
To examine how evidence on LLM persuasion has evolved over time and to investigate possible effects of how LLM models evolve (``model drift''), we conducted a cumulative meta-analysis by study year (see Figure~\ref{fig:cumulative_evolution}). In this procedure, studies are chronologically ordered and then added one by one to the pooled estimate, so that each step updates the overall effect size as if the evidence base had ended at that year. The analysis shows noticeable shifts in effect size estimates as studies accumulated from 2023 through 2025. Early studies from 2021 and 2022 tended to point toward LLMs being somewhat less effective than humans in persuasion. With the addition of more recent studies in 2024 and 2025, the cumulative estimate moved toward a null effect that appears more stable and leans slightly in the positive direction. However, given the results from the combined moderator analysis, this time trend may not necessarily reflect the evolution of LLM models, but rather shifts in how LLMs are deployed. More recent studies employ interactive conversational designs and personalization techniques that might better exploit the persuasive strengths of LLMs.

\newpage
\section{Discussion}
\label{sec:discussion}
% [Interpretation der Ergebnisse, Limitationen]

\subsection*{Key findings}

%No average effect between LLMs and humans/ challenges both optimistic and pessimistic narratives/ persuasion is likely conditional on contextual factors/shift from binary “AI vs. Human” framing toward understanding interaction dynamics, task structure, and model capabilities.
Our meta-analysis reveals that current evidence does not demonstrate a consistent difference in persuasiveness between LLMs and humans. This runs counter to overoptimistic narratives \cite{Salvi2025, Schoenegger2025}, while also challenging skeptical perspectives that downplay the persuasive capacity of LLMs \cite{Hackenburg2024, Simon2023}. Instead, the substantial heterogeneity observed across studies suggests that persuasive effectiveness is likely conditional on contextual factors. Accordingly, we argue for a conceptual shift away from binary ``AI versus human'' framings and toward a more nuanced understanding of how model capabilities, communication design, and task characteristics jointly shape persuasive effects.

%bündel heterogeneity in 1 absatz damit es nicht zu lang wird 
%The type of LLM used (e.g., GPT-3, Claude 3.7) does not explain effectiveness differences unless considered in interaction with other moderators.
%Interactive setups significantly improve LLM persuasiveness compared to one-shot formats, but only in combined model (g = −0.3520, p = .023 in combined model).

%Only the combined model explains between-study variation (R² = 99.36%), identifying Claude 3.7 and one-shot formats as significantly worse than GPT-4 in interactive settings. / highlights the importance of interaction between moderators. Single-factor tests may underestimate such effects.

To better understand which design and contextual factors influence the persuasiveness of LLMs, we examined variation across studies via a model with all combined moderators. The result that individual moderator analyses (e.g., LLM model, conversation design, domain) did not yield consistently significant effects was likely due to the small number of studies and therefore insufficient power to detect differences. However, the combined model explained a large share of the variance between studies ($R^2 = 81.93\%$) and considerably reduced unexplained heterogeneity ($I^2 = 35.51\%$). This indicates that contextual factors may play an important role in shaping persuasive outcomes. For example, when holding other factors constant, interactive setups were more persuasive than one-shot formats, GPT-4-based models outperformed Claude 3.x, and health-related topics yielded stronger effects than political domains. However, with few studies and many predictors, the combined model may overfit the data.

\subsection*{Implications}

%Practitioners should not assume LLM superiority by default but consider interaction design and model-specific performance.
%(merge into 1 paragroaj)
%High-performing LLMs (e.g., GPT-4) in interactive settings are more promising for persuasive tasks such as marketing
The findings of our meta-analysis point to important implications for practice, society, and research. For practitioners (e.g., in marketing), this suggests that the persuasive effectiveness of LLMs cannot be taken for granted but must be evaluated in light of specific use cases. Multi-turn conversational interactions may offer advantages over one-shot prompts in contexts where persuasive influence is the goal, such as marketing, political messaging, or customer engagement \cite{Schoenegger2025, Feuerriegel2024}. In such interactive contexts, LLMs could draw on strengths like the ability to personalize messages, evidence-based arguments and reason coherently \cite{Salvi2025, TimmJasper2025, Bai2025,  Schoenegger2025}.

Conversely, simpler LLMs or static one‑shot deployments may underperform and even lead to unintended outcomes. Recent surveys in computational persuasion warn that minimally contextual or one-shot approaches are especially prone to bias, adversarial manipulation, and context insensitivity, which can reduce effectiveness or backfire in persuasive use cases \cite{Bozdag2025}. Furthermore, one-shot message formats may expose limitations of LLMs seen in other domains, such as creative writing, where outputs are often less original than those of humans \cite{doshi2024, Holzner2025}.

This pattern might similarly translate to the domain, as the effectiveness of LLMs might be dependent on the topic discussed. Across the reviewed studies, human-generated messages were typically more emotionally vivid and personally engaging, whereas LLM-generated texts relied more on analytical reasoning and informational coherence \cite{Bai2025, Salvi2025, Schoenegger2025, Goldstein2024, Karinshak2023}. LLMs may thus be particularly effective in domains where fact-based reasoning and logical elaboration are central to persuasion, while they may be less effective when emotional resonance, empathy, or narrative authenticity are required.
%what sbout social implications for malicious use cases
%(move to implications)
%Ethics and responsibility in LLM-based persuasion
    %What kinds of persuasion by LLMs are ethically acceptable? Under what conditions does persuasive AI become manipulative?
    %LLMs become increasingly human-like in tone and reasoning - calls for the development of a framework to guide the use of persuasive LLMs.
From a societal and ethical perspective, the findings raise concerns about the responsible use of persuasive LLMs \cite{Salvi2025}. The mechanisms that might contribute to persuasive effectiveness, such as the specific conversation design, can equally be leveraged to manipulate, deceive, or exploit users. This risk is particularly salient in high-stakes domains like political communication and misinformation \cite{zannettou2019} or when LLMs are used as chatbots for mental health support, where dynamic engagement may bypass existing safety frameworks by LLM developers. As LLMs increasingly approximate human-like tone and reasoning \cite{Nighojkar2025, Tang2024}, the boundary between legitimate influence and manipulation becomes increasingly difficult to delineate \cite{Osborne2025}. A further emerging area of concern is healthcare, where LLMs are increasingly explored for diagnostic reasoning, triage, and treatment recommendations \cite{spitzer2025, thirunavukarasu2023, williams2024}; in such settings, persuasive or overly confident outputs could anchor clinicians on incorrect decisions and pose direct risks to patient safety. This highlights the need for robust ethical guidelines \cite{Jobin2019, Canas2022, Ruane2019, gabriel2025} that define acceptable forms of AI-mediated persuasion. Developers and regulators should not only focus on technical capabilities, but also consider transparency of intent, user autonomy, and the design of interaction structures that prevent coercive or deceptive practices.

%This study reconciles previously mixed findings in the literature by showing that agent effectiveness is contingent on contextual factors
From a theoretical perspective, our findings inform our conceptual understanding of both human–AI collaboration and persuasion research. In human–AI collaboration theory, the observed heterogeneity of model capabilities, communication format, and domain aligns with frameworks emphasizing that humans and LLMs possess complementary strengths, suggesting that effective persuasion may arise when these distinct abilities are combined \cite{dellermann2021, vaccaro2024}. Qualitatively, synthesizing the observed studies, the observed heterogeneity aligns with the Elaboration Likelihood Model of persuasion: LLMs might show potential along the central route of persuasion, which relies on analytical processing, whereas human communicators retain strengths along the peripheral route, which depends on emotional, relational, and identity-based cues \cite{PettyCacioppo1986}. Moreover, LLMs offer novel opportunities for theory testing \cite{feuerriegel2025} by enabling systematic manipulation of message characteristics in controlled, large-scale environments. Such setups allow researchers to replicate classic persuasion experiments or explore theoretical mechanisms at a scale and level of experimental control that would be difficult to achieve with human participants alone \cite{argyle2025, aher2023, park2024}. By identifying conditional factors rather than relying on binary human-versus-AI comparisons, our study contributes to a more nuanced understanding of how, when, and why LLMs can be persuasive.

\subsection*{Limitations}

This meta-analysis reflects the current empirical landscape and is therefore subject to several limitations. First, the number of eligible studies is limited (i.e., $n = 7$ studies with $12$ effect size estimates), which reflects the emerging state of research on LLM-based persuasion and which may constrain certain subgroup analyses. It is important to note that the small number of included studies is a deliberate outcome of our stringent inclusion criteria, which aimed to isolate the persuasive effect attributable to the LLMs themselves and exclude studies in which differences were driven by confounding factors such as software design, so that we can attribute the effects to the LLMs' capabilities. This deliberately narrow focus improves internal validity and ensures greater comparability across measures, thereby raising data quality. Still, we are cautious in the interpretation due to the small statistical power, especially for the subgroup analyses, and because the number of studies is small relative to the number of moderators, which increases the risk of overfitting and inflated explanatory power. Thus, more research will be needed in the future to develop robust empirical evidence. Second, the dataset includes studies with considerable variation in domains, outcome definitions, and sample characteristics. While this introduces heterogeneity, it also reflects the diversity of real-world use cases for LLM-based persuasion. Third, several of the included studies are not yet peer-reviewed. Given the fast pace of LLM research, we chose to include these preprints to ensure a timely and comprehensive synthesis including recent evidence. Fourth, the available evidence is overwhelmingly based on WEIRD (Western, educated, industrialized, rich, and democratic) and U.S.-centric samples, as all studies recruited participants from online convenience panels such as Prolific, Lucid, or MTurk. As such, the generalizability of our findings to non-WEIRD populations and non–U.S. contexts remains limited and should be treated with caution.

\subsection*{Recommendations for future research}

Building on the insights from our structured literature review and meta-analysis, we outline three research directions that address key gaps in the current evidence base on LLM-driven persuasion.

\emph{Research direction 1: Understudied role of personalization in LLM persuasion.}
Our structured review revealed that only one study \cite{Salvi2025} explicitly implemented personalization in LLM-based persuasion. Prior research indicates that tailoring messages to recipient characteristics, such as personality traits, moral values, or motivational orientation, can enhance persuasive outcomes in marketing, political communication, and health promotion \cite{Matz2024, matz2017, teeny2021, feng2025}.
In the human–AI interaction literature, emerging evidence suggests that AI-generated messages can reproduce these personalization benefits, even when using minimal profile data \cite{Matz2024}. However, most current LLM studies rely on generic one-size-fits-all messages, leaving open important questions about whether established personalization effects translate to LLM-mediated persuasion, and how such tailoring interacts with model capabilities, conversation design, and domain context. 
Future research could move beyond generic one-size-fits-all messages to test whether established personalization effects hold in LLM-mediated persuasion. Studies could test varying levels of tailoring and explore how results differ across domains such as political discourse, health communication, and financial advice. Such work can clarify in which contexts personalization adds value, when it does not, and how it interacts with model capabilities, domain, and interaction format.

\emph{Research direction 2: Real-world-inspired and longitudinal study designs.}
The majority of existing LLM persuasion studies relies on simplified, one-off scenarios, such as rating a single static message. While these approaches allow for strong experimental control, they fall short of capturing the complexity of real-world persuasion contexts, where influence unfolds across multiple interactions, channels, and time points \cite{Schoenegger2025}. Persuasion in marketing campaigns, political mobilization, or customer relationship management is often cumulative in the sense that it is shaped by repeated exposures and evolving audience responses \cite{schmidt2015}. Future research could therefore benefit from settings with ecological validity that better reflect these dynamics, for example, by embedding LLM-generated messages into ongoing campaigns or sustained dialogues. Longitudinal designs would enable researchers to track whether persuasive effects persist, strengthen, or decay over time, and to identify patterns such as habituation \cite{rankin2009} or reinforcement \cite{perkins1968}.

\emph{Research direction 3: Psychological mechanisms of LLM persuasion}
Our review shows that existing LLM persuasion studies almost exclusively report outcome metrics (e.g., attitude change, intention, compliance rate) without measuring the underlying psychological processes that lead to these effects (e.g., as in \cite{Salvi2025}). This represents an important gap, as persuasion research suggests that different underlying mechanisms can produce similar observable outcomes, yet may differ in their durability, applicability across contexts, and potential for unintended effects \cite{PettyCacioppo1986}. Future studies should therefore integrate psychological measures into LLM persuasion experiments. This would make it possible to determine, for instance, whether possible advantages of personalization or interactive setups stem from deeper message processing, higher credibility, or increased engagement. Mapping these mechanisms would not only clarify why certain combinations of model and conversation design work but also help predict when LLM persuasion is likely to fail or have unintended consequences. Thus, analyzing mechanisms enables research to move beyond model-specific performance, yielding empirical evidence that remains valuable despite rapid technical improvements.

\newpage
\section{Conclusion}
\label{sec:conclusion}

%1 short parageoah

%LLMs and humans show similar persuasive effectiveness

%Moderator Analysis suggests that the effectiveness of LLMs is highly context-dependent.

%As LLM capabilities evolve rapidly, future studies should re-evaluate persuasive performance continuously.

This meta-analysis finds insufficient evidence for an overall difference in persuasiveness between LLMs and humans. However, we found substantial heterogeneity, suggesting that persuasive success may depend strongly on contextual factors such as the domain and the degree of interactivity in the communication setup. As LLMs continue to advance, future research should revisit their persuasive performance, taking into account evolving model capabilities and deployment environments. In particular, future studies would benefit from larger and more diverse samples beyond WEIRD populations, systematic testing of personalization and interactivity, and longitudinal, real-world research designs that capture how persuasion unfolds over time.

\newpage
\vspace{0.4cm}
\section*{Data availability}
The research data, code, and analysis outputs supporting the results of this manuscript are publicly available on GitHub at the following URL: \url{https://github.com/SM2982/MetaanalysisLLMPersuasion}

\newpage
\bibliography{literature}

%%%%%%%%%%%%%%%%%%%%%%%%%%%%%%%%%%%%%%%%%%%%%%%%%%%%%%%%%%%%%%%%%%%%%%%%%%%%%%

\newpage

\section*{Author contributions} 

All authors contributed to conceptualization, manuscript writing, and approved the manuscript.

\vspace{0.4cm}

\section*{Funding}

SF declares funding via the Swiss National Science Foundation (SNSF), Grants 197485 and 186932.

%We thank our partner company for their collaboration in the implementation of the field experiment. We further thank ... for their help.

\vspace{0.4cm}
\section*{Competing interests}
The authors declare no competing interests.

\newpage

\appendix
\section*{Appendix}
\subsection*{Bias}

\begin{figure}[ht]
  \centering
  \includegraphics[width=0.6\linewidth]{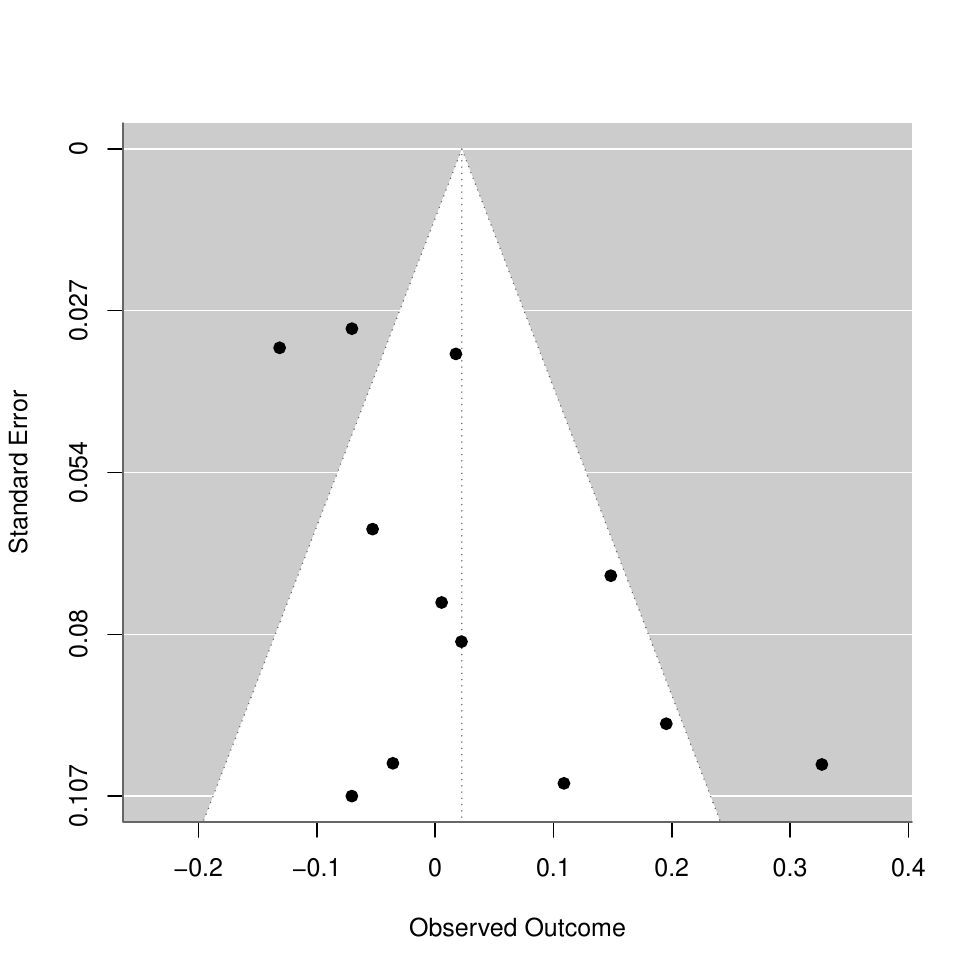}
  \caption{Funnel plot of effect sizes used to assess publication bias. No substantial asymmetry is observed.}
  \label{fig:appendix_funnel}
\end{figure}

We conducted multiple checks to assess potential publication bias and the robustness of our findings. The regression test for funnel plot asymmetry was significant (Egger's regression) ($p = .018$), indicating potential small-study effects.  However, the trim-and-fill analysis did not impute any missing studies on either side of the funnel. The corresponding funnel plot appeared mostly symmetrical except for one small-sample study that showed a notably high positive effect, likely explained by its use of personalization \cite{Salvi2025}  (see Figure~\ref{fig:appendix_funnel}). Overall, the significant Egger’s test likely reflects the small sample size and substantial heterogeneity, whereas the trim-and-fill and funnel plot analyses indicate a low risk of publication bias.

\begin{figure}[ht]
  \centering
  \includegraphics[width=0.7\linewidth]{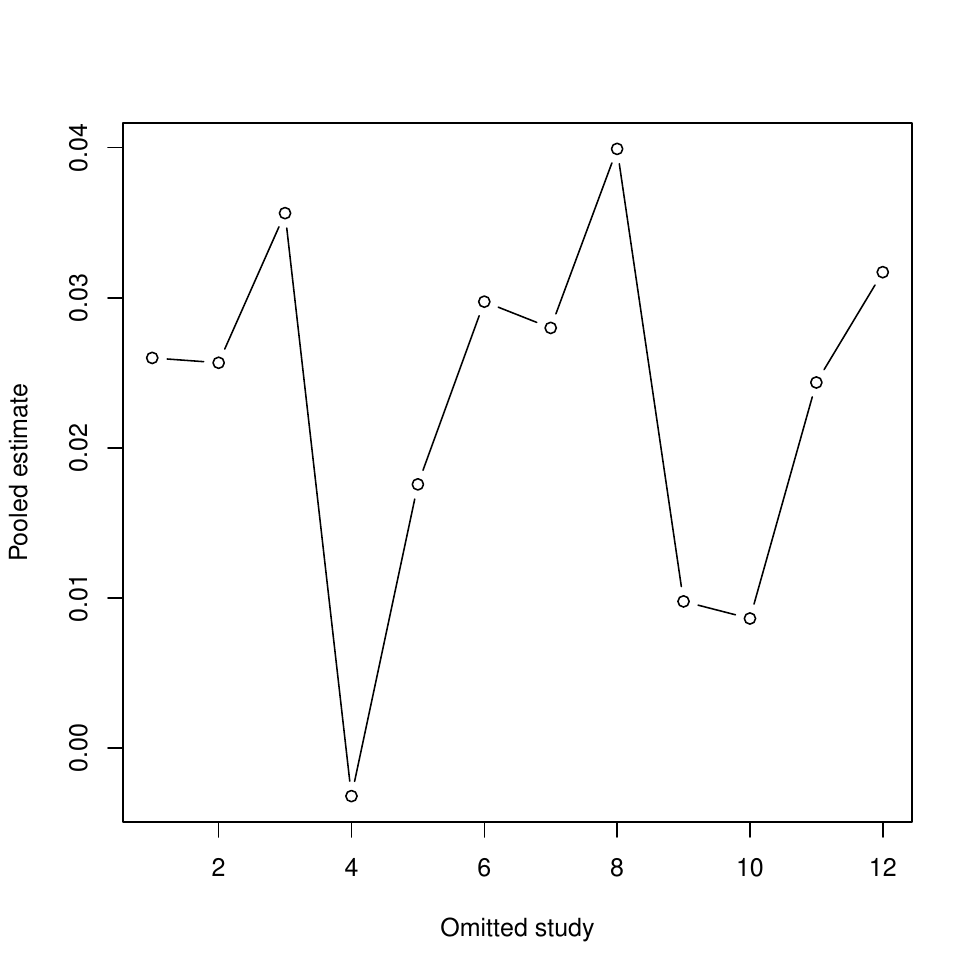}
  \caption{Leave-one-out sensitivity analysis. The pooled effect size remains stable when omitting each study.}
  \label{fig:appendix_loo}
\end{figure}

Leave-one-out analyses revealed that the overall effect remained largely stable when omitting individual studies. However, one study showed noticeable influence, as its exclusion led to a downward shift of the pooled estimate (see Figure~\ref{fig:appendix_loo}). While this indicates a degree of heterogeneity, the overall inference remains robust, as the effect remains statistically non-significant in all cases.

\subsection*{Robustness}

\begin{table}[htbp]
\centering
\begin{tabular}{lccc}
\hline
\textbf{Subset of Studies} & \textbf{Effect Size (Hedges' $g$)} & \textbf{95\% CI} & \textbf{I$^2$ (\%)} \\
\hline
All studies ($k = 12$) & 0.023 & [-0.048, 0.099] & 75.97 \\
Peer-reviewed only ($k = 9$) & 0.046 & [-0.052, 0.145] & 74.47 \\
\hline
\end{tabular}
\caption{Comparison of meta-analytic results including all studies and peer-reviewed studies only. Results are consistent between all studies and peer-reviewed only.}
\label{tab:peer_review_comparison}
\end{table}

%Bias diagnostics indicate that the results are robust and not driven by individual outliers or publication bias.

%Influence diagnostics and leave-one-out analyses showed no dominant study.

%Egger's test did not indicate funnel plot asymmetry.

%Trim-and-fill analysis did not impute any missing studies (k₀ = 0).

%Either show respective plots here or in Github

\end{document}